\documentclass[superscriptaddress,amsmath,amssymb,aps,prl,reprint]{revtex4-1}

\usepackage{graphicx}
\usepackage{dcolumn}
\usepackage{bm}
\usepackage{hyperref}
\usepackage{notes2bib}
\usepackage{siunitx}
\usepackage{amsmath}
\usepackage[usenames,dvipsnames]{xcolor}
\usepackage[utf8]{inputenc}

\setcitestyle{super}

\begin{document}

\preprint{APS/123-QED}

\title{On-chip detection of anisotropic thermopolarization in quartz}

\author{Shuichi Iwakiri}
\email{iwakiri.shuichi@nims.go.jp}
\affiliation{
 National Institute for Materials Science, Tsukuba, Japan
}

\author{Yasumitsu Miyata}
\affiliation{
 National Institute for Materials Science, Tsukuba, Japan
}

\author{Takao Mori}
\affiliation{
 National Institute for Materials Science, Tsukuba, Japan
}

\date{\today}
\begin{abstract}
Temperature gradients are widely used to drive and probe transport phenomena in solids, forming the basis of heat-to-charge conversion processes. 
In typical experiments, local heating is introduced to generate a temperature gradient, and the resulting electrical response is detected by separate electrodes. 
Such measurements usually regard heating purely as a source of thermal excitation.
Here, we show that heating inherently generates mechanical stress through thermal expansion, which in turn produces measurable electrical signals via electromechanical coupling. 
Using quartz as a model piezoelectric system, we demonstrate that heat can be converted to electrical currents via thermally generated stress.
The on-chip device used in our experiment enables us to probe the anisotropy of the piezoelectric tensor through the thermally generated current, exhibiting twofold and threefold responses for X-cut and Z-cut crystals, respectively. 
We further show that the response can be detected in both current and voltage modes.
These results reveal a thermomechanical pathway for heat-to-charge conversion and establish a general platform for electrically probing thermomechanical responses in insulating materials.
\end{abstract}

\maketitle
\section{Introduction}
Temperature gradients provide the fundamental driving force for heat-to-charge conversion in solids.
In conductors, they generate electrical responses through thermoelectric effects such as the Seebeck effect \cite{AshcroftMermin1976}, the Nernst effect \cite{VonEttingshausenNernst1886}, and the nonlinear Seebeck effect \cite{Arisawa2024NonlinearThermoelectric}, as well as spin-current generation in the spin Seebeck effect \cite{Uchida2008SpinSeebeck}. 
To probe such phenomena, microfabricated devices consisting of a heater and a detector electrode, hereafter referred to as heater--detector devices, are widely employed. 
In these devices, a heater generates a temperature gradient via Joule heating \cite{Fangohr2011}, while a separate detector electrode detects electrical responses. Within this framework, the heater is typically regarded solely as a source of temperature gradients.

An inevitable consequence of heating, however, is its coupling to mechanics. 
Thermal expansion, a macroscopic manifestation of phonon anharmonicity, generates strain and stress in solids, which can couple to electrical properties through electromechanical effects such as piezoelectricity \cite{Mason1950_piezoelectric_crystals}. 
In piezoelectric materials, the polarization $\mathbf{P}$ is linearly related to the stress  $\boldsymbol{\sigma}$ through the piezoelectric tensor $\mathbf{d}$ as
$\mathbf{P} = \mathbf{d} \, \boldsymbol{\sigma}$, providing a direct mechanism by which thermally generated stress is converted into electrical polarization.
In bulk crystals, such coupling is known as secondary pyroelectricity, where thermally induced strain modifies polarization via piezoelectric interactions \cite{Liu2018_mechanisms_pyroelectricity,Masuki2023_quasiharmonic_pyroelectricity_PRB,Liu2016_firstprinciples_pyroelectricity}. 
A representative example is a piezoelectric crystal such as quartz, in which thermally generated stress is converted into polarization through the piezoelectric tensor \cite{Jain2023_pyroelectric_JPCM,Liu2018_mechanisms_pyroelectricity_PRL,Masuki2023_quasiharmonic_pyroelectricity_PRB,Poplavko1998_quartz_sensors,Poplavko2025_piezoelectric_intrinsic_polarity}. 
While previous studies have primarily relied on bulk measurement setups designed for such effects, this connection suggests that widely used heater--detector devices may also inherently generate and detect thermomechanically induced electrical responses, providing a route to electrically probe insulating materials within a simple device architecture.

In this work, we demonstrate that heater--detector devices can directly detect thermomechanically generated electrical signals, revealing an intrinsic but previously overlooked functionality of this widely used geometry. 
Using X-cut and Z-cut quartz, we show that the observed signal follows the crystal orientation dependence expected from piezoelectric tensor. 
The symmetry is further confirmed by analytical evaluation and finite-element simulations. 
We also demonstrate that the signal can be detected as both current and voltage across different electrode configurations.

These results establish an on-chip platform for generating thermally induced stress and electrically probing responses in insulating materials. 
The underlying electromechanical response is not limited to piezoelectricity but can be extended to more general couplings such as flexoelectricity. 
In flexoelectricity, the polarization $P_i$ is generated by strain gradients $\partial \varepsilon_{jk} / \partial x_l$, expressed as $P_i = \mu_{ijkl} \partial \varepsilon_{jk} / \partial x_l$ ($i,j,k,l = x,y,z$) with the flexoelectric tensor $\mu_{ijkl}$. We discuss the thermopolarization generated by this mechanism in our accompanying paper \cite{Iwakiri_joint_submission}.

\begin{figure*} \begin{center} \includegraphics[scale=0.28]{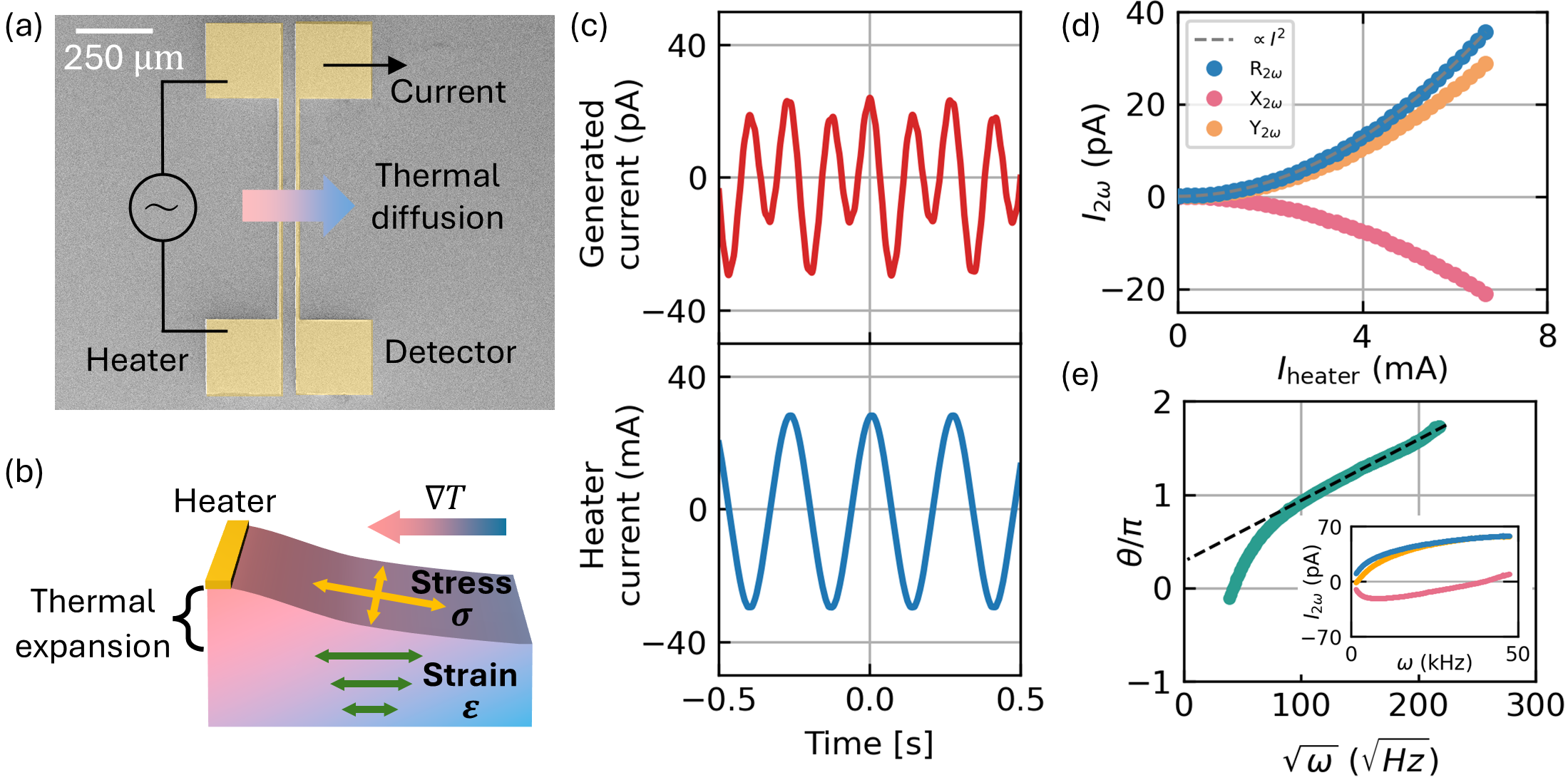} \caption{(a) Image of the device. (b) Schematic of the thermomechanical effects created by an on-chip heater. Thermal stress, strain, and expansion are illustrated. (c) Time-dependent measurement of the generated current measured at the heater (top), compared with the current applied to the heater (bottom). (d) Lock-in detection of the generated current. (e) Lock-in phase of the generated current. The inset shows the frequency dependence of the amplitude. } \label{concept} \end{center} \end{figure*}

\section{Device structure}
We begin by studying X-cut $\alpha$-quartz, which possesses a piezoelectric response that converts in-plane stress into out-of-plane polarization \cite{Labeguerie_Quartz_2010}. 
To investigate thermomechanical current generation, we fabricate an on-chip heater and a metallic detection pad onto the quartz substrate, as shown in Fig.~\ref{concept}(a). 
The heater and the detector patterns are made by maskless photolithography (MicroWriter ML3 Pro, Durham Magneto Optics Ltd.) using AZ-5214E photoresist, baked for 1 minute at 100 $^\circ$C. Both patterns have a width $10~\mu$m and a length $650~\mu$m separated by $40~\mu$m away from each other. Probing pads with $250~\mu$m $\times$ $250~\mu$m are attached. After developing the pattern, a 50 nm-thick Au film is formed by RF sputtering, and the resist is carefully lifted off in acetone.

The heater is driven by an alternating current $I = I_{\mathrm{heater}}\sin\omega t$, which produces periodic Joule heating and generates temperature gradients.
Under such a thermal gradient, the substrate undergoes space-dependent thermal expansion, which generates lateral stress as illustrated in Fig. \ref{concept}(b).
These thermomechanical fields provide the driving forces for polarization through piezoelectricity in quartz.
In the following measurements, an AC excitation voltage $V_{0}=0.004$--$2~\mathrm{V}$ at frequency $\omega/2\pi=0.01$--$50~\mathrm{kHz}$ is applied to the heater (resistance $R\simeq 150~\Omega$). The excitation current is $I_{\mathrm{heater}} = V_0/R$. The detection pad is connected to a transimpedance amplifier (Femto DLPCA-150, FEMTO Messtechnik GmbH), and the resulting current is fed to an oscilloscope or a lock-in amplifier (SRS810, Stanford Research Systems Inc.) with a time constant of 1 sec. We detect the second-harmonic voltage at $2\omega$ and convert it to the alternating current $I_{2\omega}$ using the amplifier gain ($10^{6} - 10^{8}~\mathrm{V/A}$).

\section{Observation of thermomechanical current generation}
The thermomechanical-to-electrical conversion yields a current
$I \propto \frac{dP}{dt}
= \frac{\partial P}{\partial \sigma}
  \frac{\partial \sigma}{\partial T}
  \frac{\partial T}{\partial t}$,
where $P$, $t$, $\sigma$, and $T$ denote polarization, time, stress, and temperature, respectively. 
Here, $\partial P/\partial \sigma$ represents the piezoelectric coupling, $\partial \sigma/\partial T$ corresponds to the thermally induced stress, and $dT/dt$ arises from the time-dependent heating of the device.

Since the heater temperature follows
$T(t) \propto I_{\mathrm{heater}}^{2}\sin^{2}(\omega t)$, its time derivative becomes $\frac{dT}{dt} \propto I_{\mathrm{heater}}^{2}\sin(2\omega t)$, predicting an electrical response at twice the heater frequency and quadratic in the heater current.
Indeed, when the heater is driven at $3~\mathrm{Hz}$, we observe a clear current signal at $6~\mathrm{Hz}$, as shown in Fig.~\ref{concept}(c), demonstrating the heat-to-charge conversion.
In these measurements, we employ sufficiently low driving frequencies to minimize capacitive coupling between the heater and the detector electrode, which is still visible as the modulation of the detected current amplitude.

To quantify the current generation, we measure the second-harmonic current response $I_\mathrm{2\omega}$ by lock-in amplifier at 1 kHz. 
Figure \ref{concept}(d) shows that both the in-phase ($X$) and out-of-phase ($Y$) components of $I_{2\omega}$ increase quadratically with $I_{\mathrm{heater}}$, in agreement with the above observation.
We also find that the lock-in phase and its frequency dependence encode the thermal origin of the generated signal. In the device, the temperature approximately follows the one-dimensional diffusion equation $\frac{\partial T}{\partial t} = D\,\frac{\partial^{2}T}{\partial x^{2}}$ where $D$ is the thermal diffusivity. 
The solution of the diffusion equation with a sinusoidal heating at $x=0$ has the following form,
\begin{equation}
T(x,t)=T_0e^{-kx}e^{i(\omega t-kx-\frac{\pi}{4})}.
\end{equation}
Here, the phase lag $k=\sqrt{\frac{\omega}{2D}}$ is the hallmark of the diffusive transport \cite{Angstrom1863_thermal_conductivity, Morikawa1998_high_order_harmonics, Morikawa2009_polyimide_diffusivity,Ordonez-Miranda2023_Analytical3omega}. 
It is important to note that the phase is determined solely by the thermal diffusivity $D$ and experimentally defined parameters.
The generated current $I_{2\omega}\propto dT/dt$ inherits the same frequency dependence of the phase.

As shown in Fig. \ref{concept}(e), we find the phase $\theta(\omega)$ becomes proportional to $\sqrt{\omega}$ at high frequency. 
From the fitting, we find that the thermal diffusion coefficient of the X-cut quartz to be $\sim (7\pm1) \times10^{-6} \mathrm{m^2/sec}$, which agrees well with the literature value \cite{Breuer2021_quartz_thermal_transport}.
The deviation of the data from the fit is due to the breakdown of the one-dimensional thermal diffusion model at low frequency, where the thermal diffusion length is long and heat diffusion into the stage is not negligible \cite{Morikawa1998_high_order_harmonics, Morikawa2009_polyimide_diffusivity}.
This agreement between the experiment and the thermal diffusion model demonstrates that the generated current is of thermal origin.
\section{Origin of the generated current}

\subsection{Finite element method simulation}
To gain insight into the origin of the stress and the role of temperature gradient, finite-element calculations are carried out using the coupled temperature--displacement step function of Prepomax v2.4.0 software. This enables the simultaneous simulation of the temperature field and the mechanical displacements.

The simulation employs a rectangular isotropic solid substrate approximating Z-cut quartz with dimensions of $5~\mathrm{mm} \times 5~\mathrm{mm} \times 0.5~\mathrm{mm}$.
The bottom surface of the substrate is mechanically and thermally fixed in order to emulate a rigid supporting stage at 20 $^\circ$C, while no other mechanical constraints are introduced. In the simulations, neither the heater nor the detector electrodes is included explicitly. Instead, the region corresponding to the location of the heater is represented as a localized thermal input volume.
The material parameters used in the analysis are as follows.
For simplicity, we neglect the anisotropy of physical parameters of quartz.
The linear thermal expansion coefficient is $\alpha = 1.33\times10^{-5}~\mathrm{K^{-1}}$, 
the thermal conductivity is $\kappa = 14.8~\mathrm{W\,m^{-1}\,K^{-1}}$, 
and the specific heat capacity is $c_\textrm{p} = 787~\mathrm{J\,kg^{-1}\,K^{-1}}$. 
The elastic properties are characterized by a Young's modulus of $E = 7.6\times10^{10}~\mathrm{Pa}$ 
and a Poisson's ratio of $\nu = 0.17$. 
The mass density is $\rho = 2.65\times10^{3}~\mathrm{kg\,m^{-3}}$.
The environmental temperature is set to $20^\circ\mathrm{C}$. For simplicity, the solution is obtained under the steady-state condition.
The total Joule heating is calculated from a typical heater resistance of $150~\Omega$ driven by a current amplitude of $4~\mathrm{mA}$. The geometric volume of the heater region, $650~\mu\mathrm{m} \times 10~\mu\mathrm{m} \times 100~\mathrm{nm}$, is used to estimate the nominal power, which amounts to approximately $3700 ~\mathrm{W\,mm^{-3}}$.
\begin{figure}
\begin{center}
\includegraphics[width=\columnwidth]{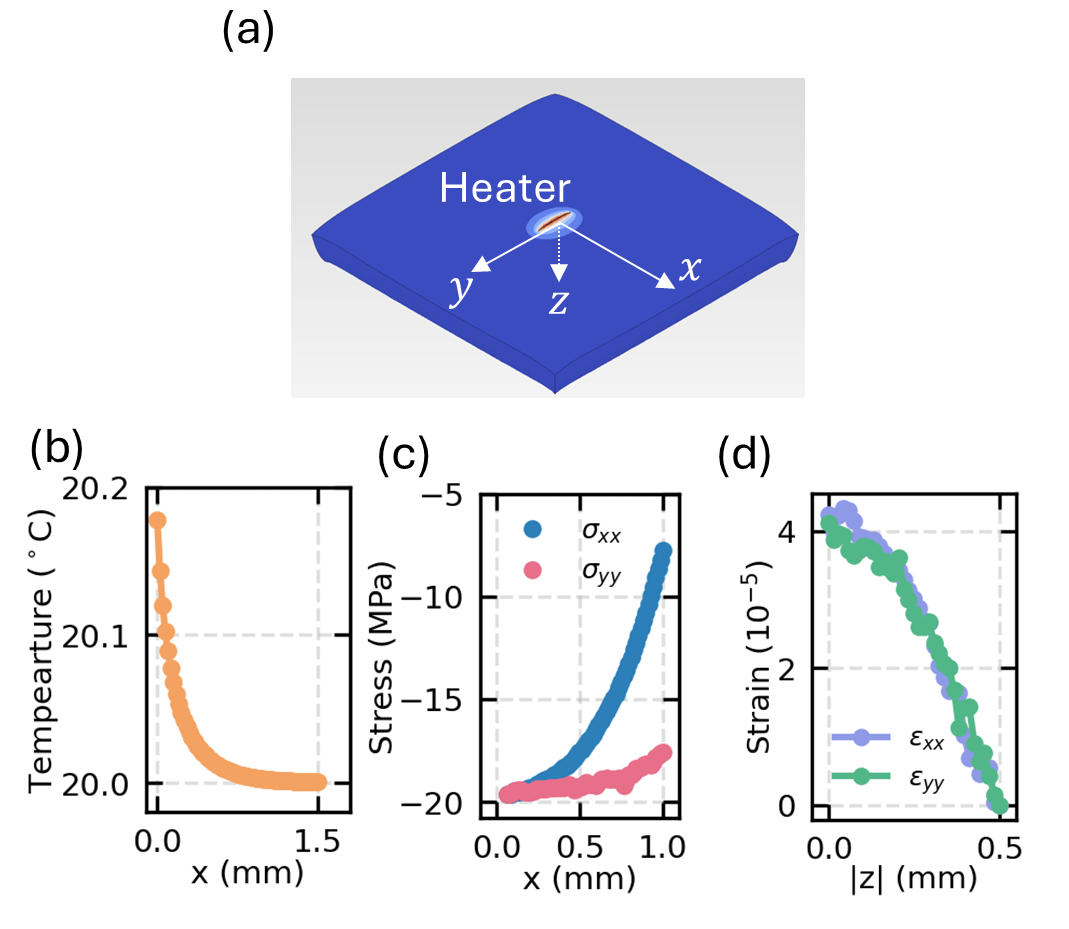}
\caption{(a) Schematic of the FEM model used in the calculations. (b) Calculated temperature profile along $x$ axis. (c) In-plane stress along $x$ axis. (d) In-plane strain along the $z$ axis.}
\label{FEM_elec}
\end{center}
\end{figure}

The simulations show that the temperature increase is concentrated near the heater position and relaxes gradually with distance, as shown in Fig. \ref{FEM_elec}(a) and (b). At the heater location, the substrate temperature increases by less than a Kelvin, while it returns close to the ambient value within a few hundred micrometers. 

Next, the in-plane thermal stresses are analyzed at the same location. The FEM output shows that the stress components $\sigma_{xx}$ (parallel to the heat flow) and $\sigma_{yy}$ (perpendicular to the heat flow) are both large near the heater, but relax as the distance from the heater increases (see Fig. \ref{FEM_elec}(c)).
Note that the magnitude of the strain $\sigma_{xx}$ and $\sigma_{yy}$ are similar at $x\sim0$, while their gradients are significantly different. This relation plays a role in the crystal orientation dependence as we discuss later.

Finally, the strain distribution normal to the substrate surface is evaluated at the detector located $40~\mu\mathrm{m}$ away from the heater region. The strain components $\varepsilon_{xx}$, parallel to the in-plane heat-flow direction from the heater, and $\varepsilon_{yy}$, along the long axis of the heater and perpendicular to the heat flow, are computed (see Fig. \ref{FEM_elec}(d)). Both components exhibit depth-dependent profiles across the substrate thickness, indicating the presence of out-of-plane strain gradients. Such a strain gradient itself does not play a role in piezoelectricity, but contributes to polarization through flexoelectricity in general \cite{Iwakiri_joint_submission}.

These simulations, although simplified, provide consistent evidence that heating leads to the emergence of spatially-dependent strains, in-plane stresses, and out-of-plane strain gradients in the substrate. 

\subsection{Phenomenology for the induced charge}
\begin{figure}
\begin{center}
\includegraphics[scale=0.32]{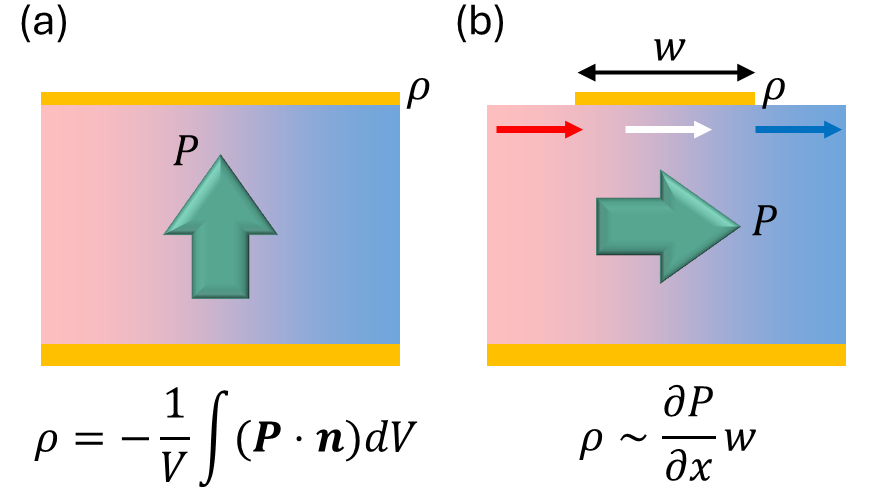}
\caption{Models of the surface charge density induced by polarization for (a) a perpendicular polarization with parallel-plate electrodes and (b) an in-plane polarization with a finite-width electrode.
}
\label{chargemodel}
\end{center}
\end{figure}

Here, we discuss how the polarization generated inside the insulator is converted into the electric current measured at the metallic detector electrode.

First, we consider the easiest phenomenology where the polarization $\mathbf{P}$ is perpendicular and the electrodes are infinitely large parallel plates (Fig. \ref{chargemodel}(a)). Kretschmer and Binder derived a surface charge density $\rho$ induced in such a case \cite{Kretschmer1979_surface_effects_ferroelectrics}
\begin{equation}
\rho=-\frac{1}{V}\int_V \mathbf{P}\cdot\mathbf{n}\, dV.
\end{equation}
This result is generic for a distributed polarization $\mathbf{P}(\mathbf{r},t)$. As long as the fringing field is neglected, this formula can be used for electrodes with finite width.

In the present geometry, however, the polarization orients predominantly in-plane (Fig. \ref{chargemodel}(b)) as we will see in the following section. Also, the electrode has a finite width $w$ and the length is much larger than it.
We consider a phenomenology for such a case by approximating the problem as one-dimensional along in-plane axis $x$.
Noting that the polarization can be understood as a pair of positive and negative charges \cite{Landauer1957_domain_formation_BaTiO3}, we can assume that only the polarization vectors at the edge of the electrode (red and blue arrows in Fig. \ref{chargemodel}(b)) contribute to the surface charge, while the one completely beneath the electrode (white arrow) does not contribute to the charge because the image charge created by the negative and positive end of the polarization is canceled.
Then, the induced charge is roughly proportional to the difference in polarization across the electrode edges,
\begin{equation}
\rho \propto \, \frac{\partial P_{x}}{\partial x} \propto P_{x}(x+w) - P_{x}(x),
\end{equation}
where $w$ is the electrode width.
Thus, the electrode effectively detects the spatial variation of the polarization rather than its magnitude.
A time-dependent polarization gradient produces a measurable current,
\begin{equation}
I \propto \frac{d\rho}{dt}
\propto \frac{\partial^2P}{\partial t\partial x}.
\label{eq:currentgeneration}
\end{equation}

\section{Decomposing the stress field}
\begin{figure*}
\begin{center}
\includegraphics[scale=0.3]{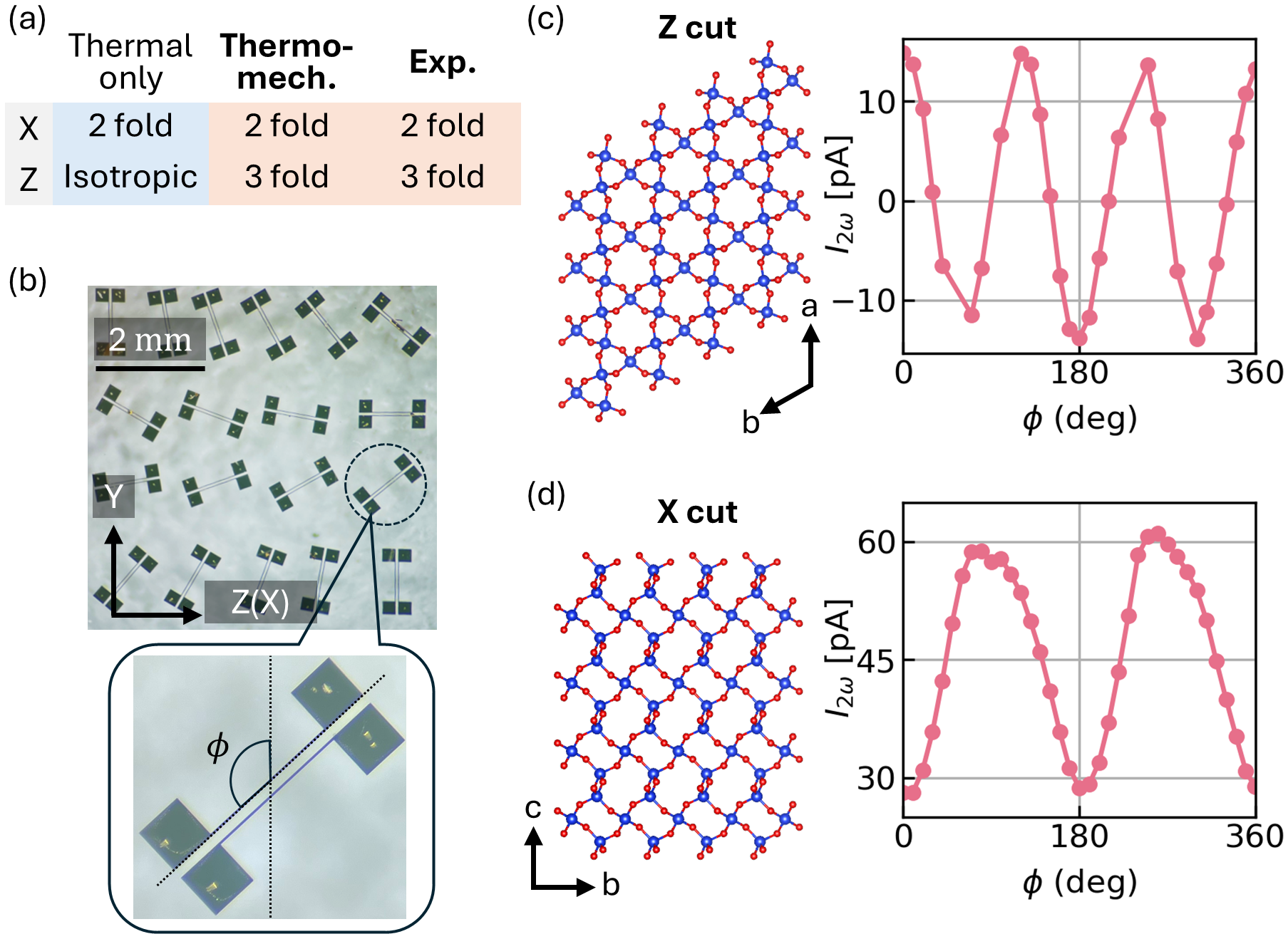}
\caption{(a) Expected symmetry in X and Z cut quartz for a purely thermal signal, thermo-mechanical signal, compared with the one observed in our experiment. (b) Picture of the device array used in the angular dependence measurement. (c,d) Angular dependence of the quadratic component of the generated current in Z cut and X cut quartz.
}
\label{symmetry}
\end{center}
\end{figure*}

The device architecture provides a direct on-chip probe of thermomechanically generated stress.
By exploiting the anisotropy of the quartz piezoelectric tensor, the crystal-orientation dependence of the measured current allows the stress components generated by the heater to be identified.

As shown in Fig. \ref{symmetry}(a), considering the crystal symmetry of X and Z cut quartz, one expects distinct angular responses for different wafer cuts.
Z-cut quartz is nearly isotropic in its in-plane thermal conductivity, whereas X-cut quartz exhibits a twofold anisotropy.
Moreover, the piezoelectric tensor produces a threefold symmetry for Z-cut quartz and a twofold symmetry for X-cut quartz.
Therefore, the angular dependence of the generated current allows us to distinguish whether the signal originates from thermomechanical coupling or from thermal artifacts.

To probe this behavior, we fabricate an array of devices with the heater–detector axis tilted by an angle $\phi$ relative to the crystal axes, as shown in Fig.~\ref{symmetry}(b).
The measured responses show the expected symmetry signatures:
X-cut quartz exhibits a clear twofold modulation of $I_{2\omega}$ [Fig.~\ref{symmetry}(c)], whereas Z-cut quartz exhibits a robust threefold modulation [Fig.~\ref{symmetry}(d)].
These observations indicate that the measured current originates from thermomechanically generated stress acting through the piezoelectric tensor.

We next find that the observed angular dependence is the fingerprint of the stess fields in the device.
In quartz, the thermal stress couples to the polarization or the displacement field according to the following relation.

The piezoelectric tensor $\mathbf{d}$ of $\alpha$-quartz \cite{Labeguerie_Quartz_2010} is
\begin{equation}
\begin{bmatrix}
P_x \\ P_y \\ P_z
\end{bmatrix}
=
\begin{bmatrix}
d_{11} & -d_{11} & 0 & d_{14} & 0 & 0\\
0 & 0 & 0 & 0 & -d_{14} & -2d_{11}\\
0 & 0 & 0 & 0 & 0 & 0
\end{bmatrix}
\begin{bmatrix}
\sigma_{xx}\\ \sigma_{yy}\\ \sigma_{zz}\\ \tau_{yz}\\ \tau_{xz}\\ \tau_{xy}
\end{bmatrix},
\label{eq:piezo-voigt-rev}
\end{equation}
where $P_i$ is the polarization, $\sigma_{ij}$ is the stress, $\tau_{jk}$ is the shear with $i,j,k=x,y,z$.

To express the polarization in the device coordinate system fixed on the device plane $(x',y',z')$, we rotate the piezoelectric relation.
Let $(x,y,z)$ denote the crystallographic axes and $(x',y',z')$ the device axes.
The two frames are related by an orthogonal rotation matrix $R$ such that
$\mathbf{e}_{i'} = R_{i'i}\mathbf{e}_i$.
The polarization vector and stress tensor transform as
$\mathbf{P'} = \mathbf{R}\,\mathbf{P}\,$ and $\mathbf{\sigma} = \mathbf{R^T}\,\mathbf{\sigma'}\,\mathbf{R}\,=\mathbf{T}\mathbf{\sigma'}$.
Then, the relation between the polarization and the strain after rotation can be written 
$\mathbf{P'} = \mathbf{R}\,\mathbf{d}\,\mathbf{T}\,\boldsymbol{\sigma'}$.

In the experiment, the detected signal is primarily sensitive to polarization variations across the short axis of the elongated detector wire ($650~\mu$m $\times$ $10~\mu$m), where the temperature gradient is the largest.
We therefore approximate the detected polarization as $P_{\mathrm{eff}} \simeq P_{x'},\, P_{z'}.$

Because the top surface is traction free, the stress components involving the surface normal are expected to be small near the surface. 
We therefore adopt the plane-stress approximation
\begin{equation}
\sigma_{z'z'} \approx 0,\qquad
\tau_{x'z'} \approx 0,\qquad
\tau_{y'z'} \approx 0 .
\label{eq:plane_stress}
\end{equation}

\subsection{Z-cut quartz}
For Z-cut quartz, the wafer normal coincides with the crystal $z$ axis, and the device rotation corresponds to a rotation about $z$ by angle $\phi$:
\begin{equation}
R_z(\phi)=
\begin{pmatrix}
\cos\phi & -\sin\phi & 0\\
\sin\phi & \cos\phi & 0\\
0 & 0 & 1
\end{pmatrix}.
\end{equation}


Carrying out the tensor rotation yields
\begin{equation}
P_{x'} =
d_{11}(\sigma_{x'x'}-\sigma_{y'y'})\cos 3\phi
+2d_{11}\tau_{x'y'}\sin 3\phi
+2d_{14}\tau_{y'z'} .
\label{eq:zcut_general}
\end{equation}
In this configuration $P_{z'}=0$ because the third row of the piezoelectric tensor vanishes.
Applying Eq.~(\ref{eq:plane_stress}) reduces Eq.~(\ref{eq:zcut_general}) to
\begin{equation}
P_{x'} \simeq
d_{11}(\sigma_{x'x'}-\sigma_{y'y'})\cos 3\phi
+2d_{11}\tau_{x'y'}\sin 3\phi .
\label{eq:zcut_plane}
\end{equation}

Experimentally the Z-cut signal exhibits an almost pure $\cos 3\phi$ dependence with negligible phase shift and zero offset as shown in Fig. \ref{symmetry}(c).
The absence of the $\sin 3\phi$ term therefore implies $\tau_{x'y'} \simeq 0$.
The polarization thus becomes
\begin{equation}
P_{\mathrm{eff}}(\phi) \simeq
d_{11}(\sigma_{x'x'}-\sigma_{y'y'})\cos 3\phi .
\label{eq:zcut_final}
\end{equation}

\subsection{X-cut quartz}
For X-cut quartz, the wafer normal coincides with the crystal $x$ axis, and the device rotation corresponds to a rotation about $x$:
\begin{equation}
R_x(\phi)=
\begin{pmatrix}
1 & 0 & 0\\
0 & \cos\phi & -\sin\phi\\
0 & \sin\phi & \cos\phi
\end{pmatrix}.
\end{equation}


The tensor rotation gives
\begin{equation}
\begin{aligned}
P_{x'} &=
d_{11}\sigma_{x'x'}
-d_{11}(\cos^2\phi\,\sigma_{y'y'}+\sin^2\phi\,\sigma_{z'z'})
\\
&\quad
+d_{14}\sin 2\phi(\sigma_{z'z'}-\sigma_{y'y'})
\\
&\quad
+\left(-d_{11}\sin 2\phi+2d_{14}\cos 2\phi\right)\tau_{y'z'} .
\end{aligned}
\label{eq:xcut_general}
\end{equation}

The rotation also produces an out-of-plane component
\begin{equation}
\begin{aligned}
P_{z'} &=
-\!\left[d_{11}\sin2\phi-d_{14}(1-\cos2\phi)\right]\tau_{x'y'}
\\
&\quad
-\!\left[d_{11}(1-\cos2\phi)+d_{14}\sin2\phi\right]\tau_{x'z'} .
\end{aligned}
\label{eq:xcut_Dz}
\end{equation}

Applying the plane-stress condition simplifies the in-plane polarization to
\begin{equation}
\begin{aligned}
P_{x'}(\phi) &\simeq
d_{11}\sigma_{x'x'}
-\frac{d_{11}}{2}\sigma_{y'y'}
-\frac{d_{11}}{2}\sigma_{y'y'}\cos 2\phi
\\
&\quad
-d_{14}\sigma_{y'y'}\sin 2\phi .
\end{aligned}
\label{eq:xcut_expanded}
\end{equation}
while Eq.~(\ref{eq:xcut_Dz}) reduces to
\begin{equation}
P_{z'} \simeq
-\left[d_{11}\sin2\phi-d_{14}(1-\cos2\phi)\right]\tau_{x'y'}.
\end{equation}

Experimentally the response is dominated by the $\cos2\phi$ term with negligible phase shift as shown in Fig. \ref{symmetry}(d). 
This leads to $\tau_{x'y'}\simeq0$, making $D_{z'} \simeq0$ and the leading contribution become
\begin{equation}
P_{\mathrm{eff}}(\phi) \simeq
\frac{d_{11}}{2}(2\sigma_{x'x'}-\sigma_{y'y'})
-\frac{d_{11}}{2}\sigma_{y'y'}\cos 2\phi .
\label{eq:xcut_final}
\end{equation}

\subsection{Stress field}
According to Eq. (\ref{eq:currentgeneration}), the measured current stems from the spatial gradient of the induced polarization, which is generated by the stress. 
Differentiating the expressions in Eqs. (\ref{eq:zcut_final}) and (\ref{eq:xcut_final}), we obtain current for Z-cut quartz
\begin{equation}
I \propto 
d_{11}\cos 3\phi \, \frac{\partial}{\partial x}(\sigma_{x'x'}-\sigma_{y'y'}),
\label{eq:zcut_final}
\end{equation}
and for X-cut quartz
\begin{equation}
I \propto 
\frac{d_{11}}{2}\frac{\partial}{\partial x}(2\sigma_{x'x'}-\sigma_{y'y'})
-\frac{d_{11}}{2}\cos 2\phi \, \frac{\partial\sigma_{y'y'}}{\partial x}.
\label{eq:xcut_final}
\end{equation}

Indeed, Fig.~\ref{FEM_elec}(c) shows that the stress gradients satisfy
\begin{equation}
\frac{\partial \sigma_{x'x'}}{\partial x'} > \frac{\partial \sigma_{y'y'}}{\partial x'}>0,
\end{equation}
which is consistent with both the observed angular dependences and the signs predicted by Eqs. (\ref{eq:zcut_final}) and (\ref{eq:xcut_final}).

The combined symmetry analysis therefore indicates that the response is dominated by in-plane stresses, whereas shear stresses are negligible.
The stress tensor can thus be approximated as
\begin{equation}
\boldsymbol{\sigma}_{\mathrm{surf}}
\simeq
\begin{pmatrix}
\sigma_{x'x'} & 0 & 0 \\
0 & \sigma_{y'y'} & 0 \\
0 & 0 & 0
\end{pmatrix}.
\end{equation}
These results confirm the illustrated stress field in Fig. \ref{concept}(b), and demonstrates that such a mechanical field-driven response can be captured in our device geometry.

\section{Voltage-mode detection of thermomechanical polarization}
\begin{figure}
\begin{center}
\includegraphics[scale=0.3]{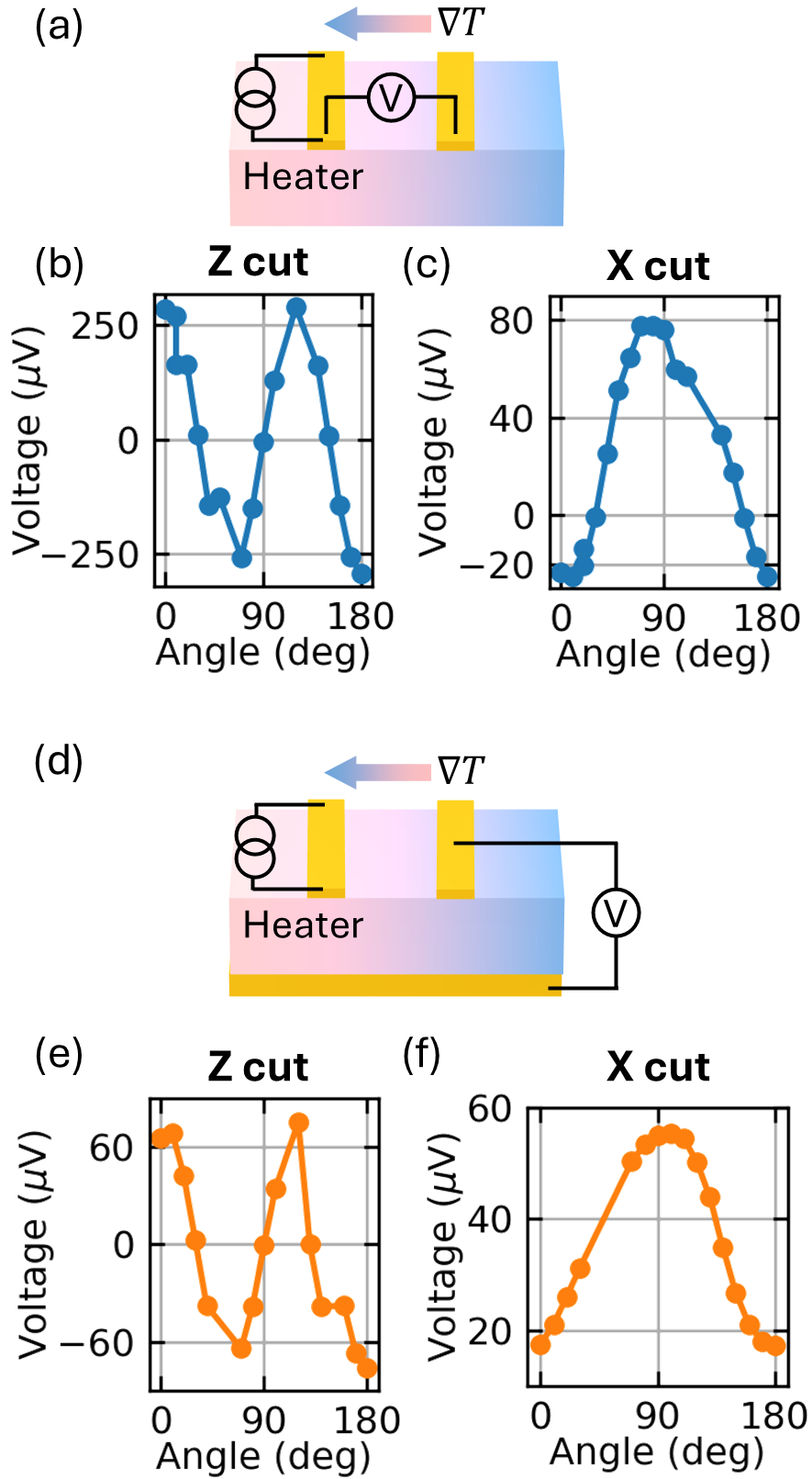}
\caption{(a) Schematic of the in-plane voltage detection. Results for (b) Z-cut and (c) X-cut samples. (d) Schematic of the out-of-plane voltage detection. Results for (e) Z-cut and (f) X-cut samples.}
\label{voltage}
\end{center}
\end{figure}
Lastly, to further validate the thermomechanically generated polarization discussed above, we examine its detection in voltage mode. 

In the present device geometry, the induced polarization is predominantly oriented in-plane. 
We therefore first measure the voltage between laterally separated heater and detector electrodes (Fig.~\ref{voltage}(a)). 
In this setup, the heater is driven by an AC voltage, and we monitor the voltages across both the heater and the detector using a lock-in amplifier operated in second-harmonic mode. The heater is excited with 3 V, which produces a current of approximately 20 mA, at a driving frequency of 100 Hz.
For convenience, we measure half of the orientation space from 0 to 180 degrees.
As shown in Fig.~\ref{voltage}(b) and (c), the measured voltage exhibits a clear threefold symmetry for Z-cut quartz, while a twofold symmetry is observed for X-cut quartz. 
These angular dependencies are consistent with those obtained from current detection (Fig.~\ref{symmetry}) and directly reflect the symmetry of the in-plane polarization vector. 

We further investigate voltage signals measured between electrodes on the top and bottom surfaces of the sample (Fig.~\ref{voltage}(d)). 
In this out-of-plane configuration, the measured voltage is influenced by fringing fields associated with in-plane polarization as well as contributions from out-of-plane polarization, leading to reduced quantitative reliability compared to the in-plane geometry. 
Nevertheless, the symmetry of the signal remains robust. As shown in Fig.~\ref{voltage}(e) and (f), threefold symmetry is observed for Z-cut quartz and twofold symmetry for X-cut quartz, consistent with the in-plane measurements.

The agreement between current-mode and voltage-mode symmetry provides evidence that the detected signal originates from thermomechanically generated polarization. 
While voltage-mode detection offers a direct probe of polarization symmetry, its magnitude depends sensitively on capacitive coupling and device geometry. 
In contrast, current-mode detection provides a more direct and quantitative measure of the generated charge.
In the accompanying paper, we employ both voltage- and current-mode detection to investigate thermopolarization in generic insulators driven by the flexoelectric effect \cite{Iwakiri_joint_submission}.


\section{Conclusion}
In summary, we have shown that an  on-chip heater--detector device inherently enables the generation and detection of thermomechanically induced electrical signals. 
Local Joule heating produces spatially nonuniform thermal expansion, which generates in-plane stress fields that couple to electromechanical responses of the material.

Using quartz as a model piezoelectric system, we have demonstrated that thermally generated stress is converted into electrical current through piezoelectric coupling. 
The observed angular dependence exhibits twofold and threefold symmetries for X-cut and Z-cut crystals, respectively, in agreement with the symmetry of the piezoelectric tensor. 
Finite-element simulations quantitatively reproduce both the stress distribution and the resulting electrical response, confirming the thermomechanical origin of the signal. 
Consistent symmetry observed in both current- and voltage-mode measurements further supports this interpretation.

These results reveal a previously overlooked functionality of heater--detector devices as a platform for electrically probing thermomechanical responses in insulating materials. 
This approach provides a route to investigating coupled thermo--mechano--electric effects in materials where conventional electronic transport measurements are not applicable.

\section{Acknowledgement}
The authors are grateful for the fruitful discussion with Dr. Hikaru Watanabe, Dr. Satoshi Ishii, and Dr. Jun Usami.
This work is partially supported by the Japan Society for the Promotion of Science (JSPS) KAKENHI Grant No. 25K21685, Murata Science and Education Foundation Grant No. M25AN110, Kanamori Foundation, and TIA Kakehashi Grant. T.M. thanks the support from JST Mirai JPMJMI19A1.

\section{Data Availability Statement}
The data that support the findings of this study are available from the corresponding authors upon reasonable request.

\bibliography{reference}

\end{document}